\documentstyle{article}
\newcommand{\ba}{\begin{eqnarray}}
\newcommand{\ea}{\end{eqnarray}}
\begin{document}
\begin{center}
{\bf A THERMODYNAMIC APPROACH TO QUANTUM MEASUREMENT AND QUANTUM
PROBABILITY} \vspace*{0.5cm}\\ Blagowest Nikolov
\\ {\it Department of Theoretical and Applied Physics, \\
Shoumen University, 9712 Shoumen, Bulgaria\\ E-mail:} b.nikolov@shu-bg.net
\end{center}
\vspace*{0.5cm}

\begin{abstract}
A simple model of quantum particle is proposed in which the particle
in a {\it macroscopic} rest frame is represented by a {\it
microscopic d}-dimensional  oscillator, {\it s=(d-1)/2} being the
spin of the particle. The state vectors are defined simply by complex
combinations of coordinates and momenta. It is argued that the
observables of the system are Hermitian forms (corresponding uniquely
to Hermitian matrices). Quantum measurements transforms the
equilibrium state obtained after preparation into a family of
equilibrium states corresponding to the critical values of the
measured observable appearing as values of a random quantity
associated with the observable. Our main assumptions state that: 
i) in the process of measurement the measured observable tends to
minimum, and ii) the mean value of every random quantity associated
with an observable in some state is proportional to the value of the
corresponding observable at the same state. This allows to obtain in
a very simple manner the Born rule.

\end{abstract}

\section{ Introduction}
In a recent paper C.Fuchs \cite{02-Fuc} has written:"Until we can explain
quantum theory's {\it essence} to a junior-high- school or high- school
student and have them walk with a deep lasting memory,we will not
understand a thing about the quantum foundations."The Born rule
 \cite{02-Sau} is the heart of quantum foundations and the aim of the present
work is to make some step in its understanding using only minimum initial
information about mechanics and probability.To do that we confine
ourselves with particles at rest and assume that the rest frame has
necessarily a {\it macroscopic} nature.This means that speaking about a
quantum particle at rest we mean that the particle accomplishes a
microscopic motion around its rest position.The simplest model of a
particle at rest is given by a classical{\it d}-dimensional linear
oscillator whose (generalized) coordinates measure the deviation of the
particle from its rest position, and whose dimension is related to the
spin of the particle.The state of the particle is understood in a purely
classical sense and described by coordinates and momenta,or equivalently,
by their complex combinations forming the state vector of the system.The
energy of the particle is equal to the energy of the particle at rest so
that the phase space trajectory is a fixed ellipsoid with microscopic
sizes comparable with the Planck scale.Thus our considerations are in
agreement with with the 't Hooft statement \cite {01-Hoo} that Planck
scale is "the most logical domain where one may expect quantum mechanics
to be replaceble by a more deterministic scenario".
 As in classical theory observables are real functions on phase space.The
 latter generate canonical transformations conserving the energy
 surface.The macroscopic smallness of the phase space coordinates allows
 to decompose observables in series retaining only the quadratic terms.It
 turns out that the latter are Hermitian forms \cite{82-Hsl,93-Ana,96-Lan}.
(In general we would come to the Weinberg theory \cite{89-Wei}.)

  The whole weirdness of quantum mechanics is hidden into the process of
 measurement.Our device measuring some observable has one input (into which
  the particles come in)  and as many outlets as the critical (i.e. minimal
  or maximal)values \cite{93-Ana} of the measured observable are.A particle entering
  into device can came out only from one of the channels having the
  corresponding critical value of the observable.This occurs at random
 with some probability.Following \cite{02-Fuc} we consider the
  probabilities in a Bayesian sense,i.e. as degrees of belief determining
  our decisions in the face of uncertainty.Then the simple (and
  reasonable) condition that one could determine the value of the
  observable making statistical measurements leads directly to the Born
  rule (cf.\cite{02-Sau}).

  \section{The model system-states and observables}

  As a model of a quantum particle at rest we consider a
  $d$-dimensional linear harmonic oscillator with mass $m$,frequency $\omega
$ and energy $E=\hbar\omega$ in accordance with de Broglie formula.Our
motivation is similar to those of Schr\"odinger\cite{30-Sch} (see also
Hestenes\cite{90-Hes}) introducing the notion of{ \it
Zitterbewegung}.Intuitively we can speak about a particle at rest only on
a {\it macroscopic } level;in fact the particle accomplishes {\it
microscopic} motion around its rest position.In our model the phase space
trajectory of this motion is given by the equation
\begin{equation}\label{1}
  \sum_{n=1}^{d}\left({p_{n}^{2}\over2m\omega}+{m\omega\over2}q_{n}^{2}\right)=\hbar
\end{equation}
where the coordinates $q_n$ measure the deviation from the rest position.
This is a $d$-dimensional ellipsoid intersecting the plane
$(q_n,p_n),n=1,2,\dots,n$ into an ellipse with area $2\pi\hbar$.When the
particle is confined to this plane we call that it has the{\it $n$-th
configuration}.(This terminology will be useful in our consideration of
measurements.)Introducing the complex coordinates
\begin{equation}\label{2}
  \psi_n=(m\omega/2)^{1/2}q_n+i(2m\omega)^{-1/2}p_n
  \end{equation}
we can rewrite Eq.(1)in the form
\begin{equation}\label{3}
  \sum_{n}|\psi_n|^{2}=\hbar
\end{equation}
The states of the particle are understood in a classical sense and
represented in matrix form by ket-vectors
\begin{equation}\label{4}
  |\psi\rangle=(\psi_1,\psi_2,\dots,\psi_d)^\top
\end{equation}
or by bra-vectors
\begin{equation}\label{5}
  \langle\psi|=(\psi_1^*,\psi_2^*,\dots,\psi_d^*)
\end{equation}
Thus our $\psi=(\psi_1,\psi_2,\dots,\psi_n)$ is not vector from  some
abstract (e.g. Hilbert) space but simply another (complex)form of the
phase space coordinates.We call it a{\it state vector},or simply a {\it
state}. In just so introduced symbols Eq.(3) looks as follows
\begin{equation}\label{6}
  \langle\psi|\psi\rangle=\hbar
\end{equation}
and can be considered as an equation of a real $2d$-dimensional sphere.

As in classical mechanics the observables are (smooth) real functions on
phase space,the latter being identified with the sphere (6).Taking in view
that macroscopically $\hbar$ is very small we shall decompose an arbitrary
observable $A(\psi,\psi^*)$ in series up to the quadratic terms: \ba &&
A(\psi,\psi^*)=A_0+\sum_{n}A_n*\psi_n+\sum_{n}A_n\psi_n^*
+\sum_{n,m}A_{nm}\psi_n^*\psi_m\\ \nonumber
&&+\sum_{n,m}B_{nm}\psi_n^*\psi_m^* +\sum_{n,m}B^*_{nm}\psi_n\psi_m \ea
Here \ba A_{nm}^*=A_{mn} \ea (since the observable is a real
function).This means that the matrix $ \hat{A}$ with elements $A_{nm}$ is
a Hermitian one.Every observable $A=A(.)$ generates a canonical
transformation which (in infinitesimal form) looks as follows \ba
\psi'=\psi-i\epsilon\partial A /\partial\psi^* \ea where $\epsilon$ is an
infinitesimal parameter and the indices are omitted.Using that
$\psi'=\psi+\epsilon\dot{\psi}$ we can write the differential equation \ba
\dot{\psi}=-i\partial A/\partial\psi^* \ea which will be called a {\it
generalized equation of motion} associated with $A$.(The dot in
$\dot{\psi}$ denotes differentiation with respect some parameter adjoint
to A;for example when $A=H$ is the energy the corresponding parameter is
time.)Now,taking in view that $A(0)=0$ (at rest our observable vanish),and
that the canonical transformations should preserve the constraint (6) we
come to the relation \ba A(\psi,\psi^*)=\sum_{n,m}A_{nm}\psi_n^*\psi_m=
\langle\psi|\hat{A}|\psi\rangle \ea Such observables will be called {\it
quantum observables}.Let us note that confining ourselves with quantum
observables only we should identify some of the vectors representing
states : vectors differing by a phase factor describe the same state.With
this in mind further we continue to call state vectors simply states.Note
that quantum observables are Hermitian forms uniquely determined by their
matrices.Moreover one can introduce a Lie algebra structure in the set of
all observables \cite {96-Lan,82-Hsl}determined by the Poisson bracket as
a Lie bracket: \ba \{A,B\}_{\psi}=i\left(\frac{\partial A}{\partial\psi}
\frac{\partial B}{\partial\psi^*}-\frac{\partial A}{\partial
\psi^*}\frac{\partial B}{\partial\psi} \right)\ea For quantum observables
$A$ and $B$ one has \ba
\{A,B\}_{\psi}=i\langle\psi|[\hat{A},\hat{B}]|\psi\rangle \ea where the
squared brackets stand for the commutator of matrices.The Hermitian
matrices form a Lie algebra with the commutator (factored by $i$) as a Lie
bracket.This equation shows that there exists an isomorphism between the
Lie algebra of quantum observables and the Lie algebra of Hermitian
matrices.This allows to identify the two algebras ,and therefore allows to
call Hermitian matrices quantum observables too.

As an example we shall cite the {\it configuration observable}
\cite{81-Gud} \ba Q=\sum_{n}n|n\rangle\langle n| \ea where
$|n\rangle=(0,\dots,0,1,0,\dots,0)^\top $, the unit being at $n$-th place.

\section{Quantum measurements}
Let us consider a particle prepared in a state $\psi$.We want to measure
the observable $A(.)$.To do that we bring it into the input of the
corresponding measuring device.Then the initial state of the particle is
destroyed ,and the particle interacting with its own environment would
became in another state.What is that state? To answer this question we
need some hypothesis about the interaction.We assume that the latter is
similar to the process of thermalization of a thermodynamic system
interacting with a reservoir (i.e. a big system).As known in such a
process the temperature of the system becomes equal to the temperature of
the reservoir.In the equilibrium thermodynamics processes like that are
described considering the behaviour of an appropriate thermodynamic
potential (energy,entropy,free energy,Gibs potential etc.):excluding
entropy (which tends to maximum)all the others tend to minimum when the
system is going to equilibrium.Guyded by this analogy we formulate our
first axiom for the quantum measurement.

{\it(QM1)In the process of measurement the measured observable tends to
minimum.}

In other words we consider the measured observable (considered as a
function on phase space) as a kind of thermodynamic potential.It is easy
to see that every critical (i.e. minimal or maximal ) value  of an
observable $A(\psi,\psi^*)=\langle\psi|\hat{A}|\psi\rangle$ on the sphere
$S=\{ \psi|\langle\psi|\psi\rangle=\hbar\}$ coincides with some eigenvalue
of the matrix $\hat{A}$.(For that it is sufficient to solve the
corresponding extremality  constraint problem.) Supposing for simplicity
that this matrix is non-degenerate we can arrange its eigenvalues as
follows: $a_1<a_2<\dots<a_d$. Denoting the corresponding eigenvectors by
$|a_n\rangle, (n=1,2,\dots,d)$ we have \ba
\hat{A}|a_n\rangle=a_n|a_n\rangle \ea From the linear algebra we know that
the eigenvectors form an orthonormal base in the linear space of all
ket-vectors:\ba\langle a_n|a_m\rangle=\delta_{nm},\sum_n|a_n\rangle\langle
a_n|=1 \ea where 1 denotes the unit matrix.Hence the spectral
decomposition follows:\ba \hat{A}=\sum_{n}a_n|a_n\rangle\langle a_n|\ea
One can prove that \cite{81-Lee} \ba a_n=\min\{A(\psi,\psi^*)|\psi\in
S_n\} \ea where \ba S_n=\{\psi\in S|\langle a_m|\psi\rangle=0,m<n\}\ea
This can be easily interpreted.Namely,if in the process of interaction
with the measuring device all states $\psi\in S$ are admissible ,the
particle is going in state $|a_1\rangle$ ,and the observable $A(.)$ takes
the minimal value $a_1$.However not all states in $S$ could be admissible
(this depends on the local environment in which the particle moves) ,and
when the set of admissible states is $S_n, (n>1)$ the particle is going in
state $|a_n\rangle$ ,and the observable takes the value $a_n$.After that
the particle come out from the device in the corresponding outlet.(To
every eigenvalue there corresponds a unique outlet.)

Obviously every such transition is a random event ,and the best what one
can do in order to make some prediction about the outcome of the
experiment, is to associate some {\it probability} to every possible
outcome.Hence it is natural to define a random quantity $A$ with values
$a_1,a_2,\dots,a_d$ .Denote the probability that $A=a_n$ by $p_n(\psi)$
when the particle (just before the measurement) is in state $\psi$.Then
the mean value of $A$ in state $\psi$ is \ba \langle A
\rangle_{\psi}=\sum_{n}a_n p_n(\psi) \ea We call the so defined random
quantity $A$ the {\it measured quantity} associated to the observable
$A(.)$ The problem now is to find the appropriate rule which would allow
us to determine these probabilities.Following Fucks \cite {02-Fuc} we
consider the probability of some outcome  in Bayesian sense ,i.e as a
degree of belief that as a result of the measurement the observable would
assume just that value.At first sight there exist many possible
assignments,and therefore some subjectivity in our decision ,but one can
suggest a natural criterion eliminating this subjectivity.This is our
second axiom:

{\it (QM1) The mean value of the measured quantity associated with some
observable in some state is proportional to the value of the observable at
that state} \ba \langle A\rangle_{\psi}=\hbar^{-1}A(\psi,\psi^*) \ea In
other words what is measured corresponds to what is in reality!Now all is
quite easy.Using Eqs.(17),(20),(21)we readily  obtain \ba
p_n(\psi)=\hbar^{-1}|\langle a_n|\psi\rangle|^{2} \ea In particular the
probability that the configuration observable Q is equal to $n$ in state
$\psi$ is \ba p_n(\psi)=\hbar^{-1}|\psi_n|^{2} \ea We obtained the Born
rule.

\end{document}